\documentclass[sn-standardnature]{sn-jnl}% referee option is meant for double line spacing
\usepackage{soul}
\soulregister{\cite}7

\jyear{2021}%

\begin{document}

\title[Article title]{Brightening of a dark monolayer semiconductor via strong light-matter coupling in a cavity}

\author[1]{\fnm{Hangyong} \sur{Shan}}
\author[2]{\fnm{Ivan} \sur{Iorsh}}
\author[1]{\fnm{Bo} \sur{Han}}
\author[3]{\fnm{Christoph} \sur{Rupprecht}}
\author[4,5,6]{\fnm{Heiko} \sur{Knopf}}
\author[4,5,6]{\fnm{Falk} \sur{Eilenberger}}
\author[1]{\fnm{Martin} \sur{Esmann}}
\author[7]{\fnm{Kentaro} \sur{Yumigeta}}
\author[8]{\fnm{Kenji} \sur{Watanabe}}
\author[9]{\fnm{Takashi} \sur{Taniguchi}}
\author[3]{\fnm{Sebastian} \sur{Klembt}}
\author[3]{\fnm{Sven} \sur{Höfling}}
\author[7]{\fnm{Sefaattin} \sur{Tongay}}
\author[1]{\fnm{Carlos} \sur{Ant\'on-Solanas}}
\author[2,10]{\fnm{Ivan} A. \sur{Shelykh}}
\author*[1]{\fnm{Christian} \sur{Schneider}}\email{christian.schneider@uni-oldenburg.de}

\affil*[1]{\orgdiv{Institute of Physics}, \orgname{Carl von Ossietzky University}, \orgaddress{\city{Oldenburg} \postcode{26129}, \country{Germany}}}

\affil[2]{\orgdiv{Faculty of Physics}, \orgname{ITMO University}, \orgaddress{ \city{Saint-Petersburg} \postcode{197101},  \country{Russia}}}

\affil[3]{\orgdiv{Technische Physik}, \orgname{Universität Würzburg}, \orgaddress{\street{Am Hubland}, \city{Würzburg} \postcode{D-97074}, \country{Germany}}}

\affil[4]{\orgdiv{Institute of Applied Physics, Abbe Center of Photonics}, \orgname{Friedrich Schiller University}, \orgaddress{\city{Jena} \postcode{07745},  \country{Germany}}}

\affil[5]{\orgname{Fraunhofer-Institute for Applied Optics and Precision Engineering IOF}, \orgaddress{\city{Jena} \postcode{07745}, \country{Germany}}}

\affil[6]{\orgname{Max Planck School of Photonics}, \orgaddress{\city{Jena} \postcode{07745},  \country{Germany}}}

\affil[7]{\orgdiv{School for Engineering of Matter, Transport, and Energy}, \orgname{Arizona State University}, \orgaddress{\city{Tempe} \postcode{85287}, \state{Arizona}, \country{USA}}}

\affil[8]{\orgdiv{Research Center for Functional Materials}, \orgname{National Institute for Materials Science}, \orgaddress{\street{1-1 Namiki}, \city{Tsukuba} \postcode{305-0044}, \country{Japan}}}

\affil[9]{\orgdiv{International Center for Materials Nanoarchitectonics}, \orgname{National Institute for Materials Science}, \orgaddress{\street{1-1 Namiki}, \city{Tsukuba} \postcode{305-0044}, \country{Japan}}}

\affil[10]{\orgdiv{Science Institute}, \orgname{University of Iceland}, \orgaddress{\street{Dunhagi 3}, \city{Reykjavik} \postcode{IS-107}, \country{Iceland}}}


\abstract{Engineering the properties of quantum materials via strong light-matter coupling is a compelling research direction with a multiplicity of modern applications. Those range from modifying charge transport in organic molecules, steering particle correlation and interactions, and even controlling chemical reactions. Here, we study the modification of the material properties via strong coupling and demonstrate an effective inversion of the excitonic band-ordering in a monolayer of WSe$_2$ with spin-forbidden, optically dark ground state.  
In our experiments, we harness the strong light-matter coupling between cavity photon and the high energy, spin-allowed bright exciton, and thus creating two bright polaritonic modes in the optical bandgap with the lower polariton mode pushed below the WSe$_2$ dark state. We demonstrate that in this regime the commonly observed luminescence quenching stemming from the fast relaxation to the dark ground state is prevented, which results in the brightening of this intrinsically dark material. We probe this effective brightening by temperature-dependent photoluminescence, and we find an excellent agreement with a theoretical model accounting for the inversion of the band ordering and phonon-assisted polariton relaxation. }




\maketitle

\section{Introduction}\label{sec1}

Atomically thin transition metal dichalcogenides (TMDCs) represent an emerging class of functional materials which are of particular interest in photonics due to their remarkable capability of efficient light emission and absorption \cite{mak2010atomically, mak2016photonics,xia2014two-dimensional}. In contrast to the vast majority of conventional semiconductors, their truly two dimensional (2D) nature results in the strong enhancement of the electron-hole Coulomb attraction, which makes their optical response to be dominated by exciton transitions even at room temperature \cite{chernikov2014exciton, wang2018colloquium, novoselov20162Dmaterials}.

Since both the valence and the conduction bands emerge from the $p$-orbitals of the transition metals, the effects of spin-orbit coupling are of central importance in excitonic ordering. Specifically, the optical selection rules allow the excitation of the lowest energy exciton in monolayers of $\rm MoSe_2$ and $\rm MoTe_2$ \cite{robert2020measurement, echeverry2016splitting}, whereas in $\rm WSe_2$ and $\rm WS_2$ it is spin-forbidden, and thus the exciton transition remains optically dark \cite{wang2015spin}. As a consequence, at low temperatures, the luminescence of these dark materials is intrinsically quenched due to the fast relaxation of excitons into the dark ground state, and needs to be thermally 'activated' \cite{zhang2015experimental}. This creates a serious obstacle for optoelectronic applications which require maximized quantum efficiency, such as light-emitting diodes \cite{mak2016photonics, withers2015light,pu2018monolayer}, and conventional \cite{ye2015monolayer, wang2012electronics, lohof2019prospects}- as well as polariton lasers \cite{anton2021bosonic, schneider2018two}.

A number of methods to enhance the luminescence efficiency of optically dark 2D materials have already been investigated. These include the band-structure shaping by spin-orbit engineering \cite{wang2015spin}, the application of an in-plane magnetic field \cite{zhang2017magnetic, molas2017brightening, robert2020measurement}, and the direct coupling of out-of-plane optical dipole moment of dark excitons with strongly localized electric fields in nanoantenna tips \cite{park2018radiative}, or a TM polarized optical mode at an oblique incidence \cite{wang2017in-plane}. However, the relevance of these approaches relying on high magnetic fields or nanoantenna tips is certainly limited for wide-spread use in applications. 

In the present work, we follow an alternative approach to reach effective brightening, which is based on the idea that the nature of the exciton ground state of the system can be qualitatively changed in the strong light-matter coupling regime \cite{shahnazaryan2019strong}, when optically-active excitons are effectively hybridized with a spatially-confined photonic mode of a microcavity, giving rise to composite elementary excitations known as exciton-polaritons  \cite{weisbuch1992observation, deng2010exciton, schneider2018two}. Attempts to engineer the emission properties of organic molecules via the inversion of the singlet-triplet state configuration by strong light-matter coupling were reported recently \cite{yu2021barrier-free, eizner2019inverting}. However, for the cases studied thus far, the detailed analyses revealed that the reverse intersystem crossing rates mostly remain invariant even as the lower polariton energy is pushed below the triplet energy \cite{eizner2019inverting}, calling for further engineering efforts. We demonstrate the phenomena of the brightening by analyzing the temperature dependence of the photoluminescence (PL) from a $\rm WSe_2$ monolayer, showing that the trend of the rapid decrease of PL intensity with decreasing temperature is inverted if a $\rm WSe_2$ monolayer is placed inside a high finesse optical cavity. We support this remarkable experimental finding by theoretical modelling of the dynamics of the mode occupancies in both regimes.

\section{Results}\label{sec2}
\subsection{Experimental geometry and brightening mechanism}\label{subsec1}

In Fig. 1a, we show the experimental system consisting of two distributed Bragg reflectors (DBRs) with a $\rm WSe_2$ monolayer flake situated in the antinode of the confined electromagnetic mode. The flake of $\rm WSe_2$ is mechanically exfoliated from a bulk crystal and it is capped with hexagonal boron nitrite (h-BN) via a deterministic dry-transfer method. Both DBRs are composed of $\rm SiO_2/TiO_2$, the corresponding Bragg wavelength is 750 nm. An optical microscope image of the sample is displayed in Fig. 1b. The $\rm WSe_2$ monolayer flake, indicated with a dashed yellow line, forms a finite-size trap for exciton-polaritons, with a size of approximately 10 x 7 $\rm \mu m^2$.  

The brightening mechanism in $\rm WSe_2$ monolayers in the strong light-matter coupling regime is illustrated in Fig. 1c. The sequence of the two lowest dark $\rm \lvert D \rangle $ and bright $\rm \lvert X \rangle $ exciton states at the $K$-point is given in the left panel. The spins of an electron and a hole in the ground state (brown line) are antiparallel to each other, and direct optical excitation of this singlet configuration is forbidden in the electric dipole approximation, making the ground state optically dark. The energy of the optically active triplet state (orange line) is $\sim40$ meV higher \cite{zhou2017probing, wang2017in-plane}.

\begin{figure}[h]%
\centering
\includegraphics[width=0.8\textwidth]{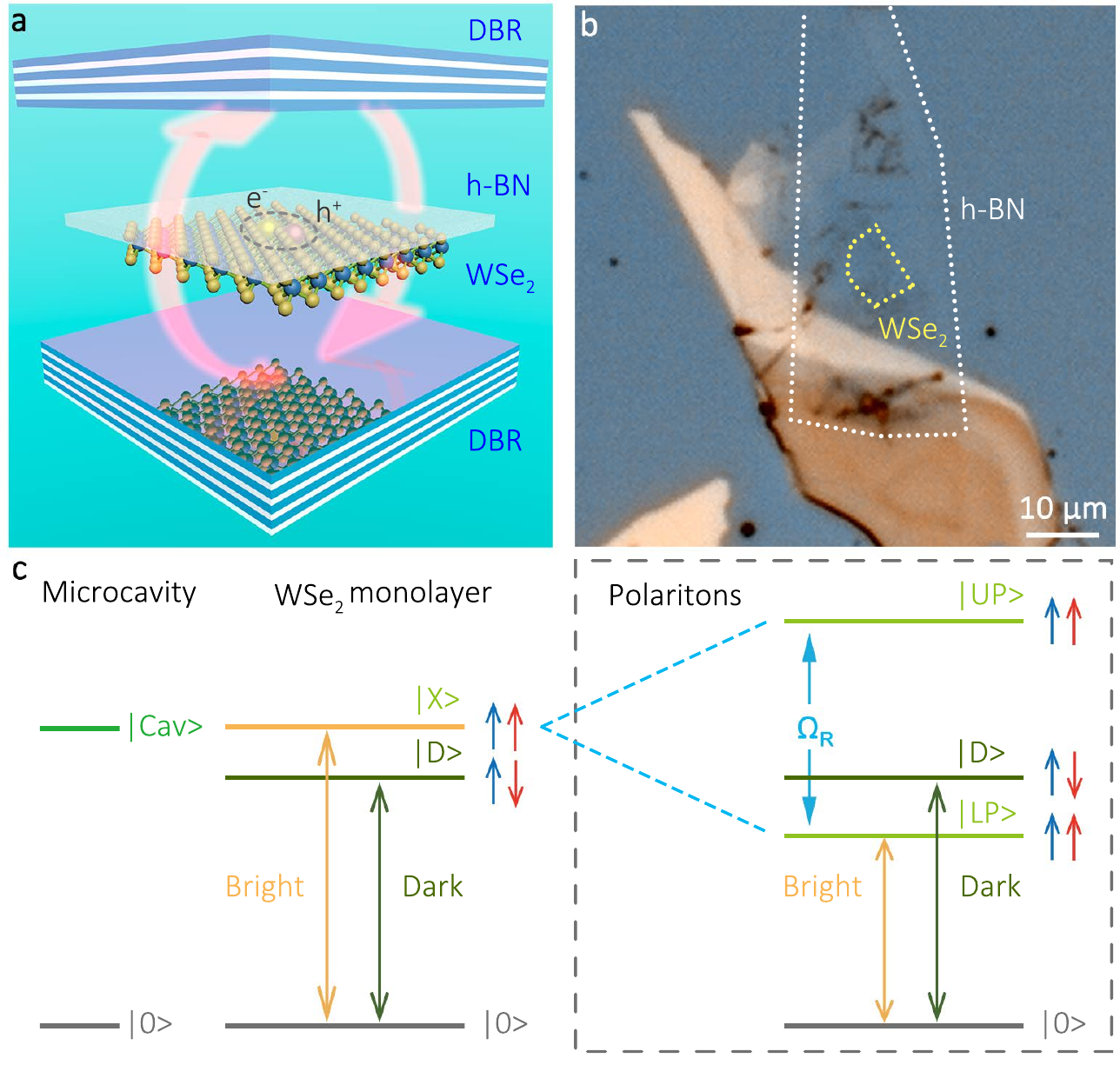}
\caption{\textbf {Sample structure and brightening mechanism.} \textbf {a} Schematic illustration of sample structure with a $\rm WSe_2$ monolayer, capped with h-BN, and embedded between two dielectric DBRs. Two arrows represent the Rabi oscillation between excitons and microcavity photons. \textbf {b} Optical microscopic image of the sample. The $\rm WSe_2$ monolayer and h-BN boundaries are indicated with yellow and white dashed lines, respectively. \textbf {c} Scheme of $\rm WSe_2$ ground state brightening via strong coupling. In pristine $\rm WSe_2$ monolayers, the optically dark exciton $\rm \lvert D \rangle $ (brown line) is the lowest transition at the $K$-point, which lies $\sim$40 meV below the bright exciton $\rm \lvert X \rangle $ (orange line). Exciton-polaritons are formed when optically bright excitons strongly couple to microcavity photons, the corresponding energy diagram is enclosed in a dashed box. The energy level of the resulting lower polaritons $\rm \lvert LP \rangle $ can be located below the dark exciton state  $\rm \lvert D \rangle $, as long as the Rabi splitting $\Omega_R$ is sufficiently large. The lower polariton branch, which inherits the spin character from the bright exciton, becomes the ground state of the coherently dressed system. Thus, the band ordering is reversed, and the intrinsically dark 2D semiconductor is effectively brightened via the strong coupling with microcavity photons.} \label{fig1}
\end{figure}

If a $\rm WSe_2$ monolayer is placed inside an optical microcavity, the strong light-matter coupling can dramatically reshape the energy spectrum of the system, as it is illustrated in the right panel of Fig. 1c. Indeed, bright excitons interact with cavity photons, giving rise to upper and lower polariton (UP and LP) modes. The Rabi splitting between these two hybrid modes $\Omega_R$, depends on the excitonic oscillator strength and cavity quality factor. If $\Omega_R$ is sufficiently large, one can push the LP energy below the dark exciton energy, reversing the energy level ordering: From one characteristic for an optically dark material, to a corresponding bright one. Note, that the LP energy can be tuned not only by changing the Rabi splitting but also by changing the relative detuning between the bare photon and exciton modes.

We use angle-resolved photoluminescence spectroscopy to study the optical properties of the coupled monolayer-microcavity system. The distribution of the PL intensity in energy and momentum is shown in Fig. 2a. The finite size of the flake results in the discretization of the energy levels of lower polaritons \cite{wurdack2021motional},which yields a dispersion-less ground state at 1.610 eV. It furthermore yields a first excited state at 1.618 eV, with emission maxima at finite k-values, and finally a continuous emission band at $\lvert k_{\parallel} \rvert$ \textgreater 2 $\rm \mu m^{-1}$. The full quantitative description of the modes in connection with the precise shape of the monolayer is derived in Ref. \cite{shan2021coherent} by accounting the energy spectrum of polaritons in the presence of an external potential V(r) via numerically solving a {Schrödinger} equation for polaritons. The slightly asymmetric distribution of PL intensity at high energy results from the irregular geometry of $\rm WSe_2$ flake, as discussed in detail in Ref. \cite{shan2021coherent}.

\begin{figure}[h]%
\centering
\includegraphics[width=0.9\textwidth]{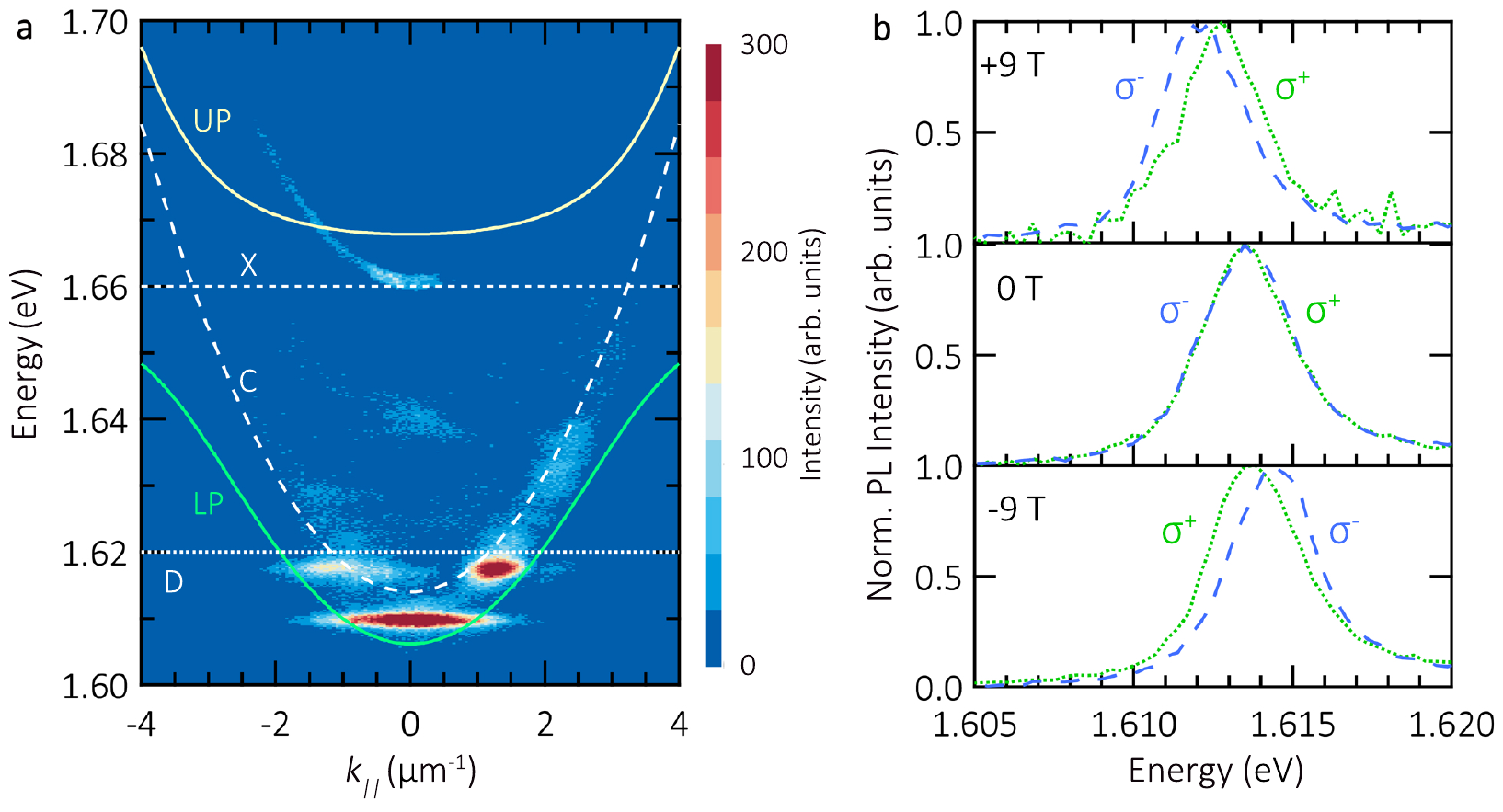}
\caption{\textbf {Polariton dispersion relation and valley-Zeeman effect.} \textbf {a} Dispersion relation of exciton-polaritons at ambient conditions. The exciton (X) and microcavity photon (C) are represented by dotted and dashed lines, respectively. The dark exciton is marked as D. The upper and lower polariton (UP and LP) branches are plotted as solid lines. The discrete energy modes of LP are a typical dispersion relation of polaritons confined at a finite-size trap. \textbf {b} Normalized circularly-polarized PL intensity spectra of  exciton-polaritons under a magnetic field at room temperature. These spectra are extracted from the dispersion relations at zero in-plane momentum $k_{\parallel}{=}0$. From top to bottom panels, the applied magnetic field is +9, 0 and -9 T, respectively. $\rm \sigma^+$ ($\rm \sigma^-$) denotes emission of light with right (left)-hand circular polarization. An energy splitting is observed under the application of magnetic fields.} \label{fig2}
\end{figure}

We notice, that the model which we apply in supplementary section A.2 is a simplified and less technical, yet analogous approach based on the discretization of the photon field as prior to the coupling to the excitons. In the model, the discretization is considered first in the photon field: Due to the lateral confinement of the cavity and the flake, the photonic modes are essentially discretized. This discretization of photon modes then translates into the discretized polariton spectrum. Hence, this model is a reduced version of the full solution to the {Schrödinger} equation. It is important to point out, that these two formulations are equivalent and yield the same polaritonic ground and first excited states.

To fit experimental data, we use the two coupled oscillator model, with parameters extracted from the experimental data. The energies of the bright exciton (X) and cavity photon (C) are 1.660 eV and 1.614 eV, respectively, the Rabi splitting is 41 meV.

 Due to the red-detuned conditions of our microcavity, and the fact that polaritons efficiently populate the ground-state of the polariton trap, a direct verification of strong coupling conditions via mapping of the Rabi-splitting is difficult. However, it is still possible to directly verify the strong coupling conditions of our TMDC-cavity system unambiguously by resorting to magnetic field measurements:  In atomically thin TMDCs, the valley pseudospin is associated with magnetic moments \cite{srivastava2015valley, aivazian2015magnetic}. Since exciton-polaritons inherit the spin from excitons, they also experience the valley-Zeeman effect, similar to bare excitons. In contrast, the effect is obsolete for cavity photons \cite{lundt2019magnetic}, rendering magnetic measurement a powerful and elegant tool to distinguish hybrid exciton-polaritons from pure photonic modes.

In Fig. 2b, we plot circularly polarized components of the PL recorded in the external magnetic field. For the sake of clarity, both components have been normalized in their intensity. 
The slight energy offset between the two panels is induced by the reduction of air pressure in our experimental apparatus which renormalizes the cavity energy (see supplementary section Fig. A2 for more details). 
One clearly sees the energy splitting of $\sim 0.5$ meV between $\rm \sigma^+$ and $\rm \sigma^-$ polarized components in the magnetic field of +9 T (top panel), which unambiguously indicates the presence of the excitonic component. As expected, the sign of the Zeeman-splitting is changed if the direction of magnetic field is inverted (-9 T, bottom panel). It is worth noting, that the effect persists for weak pump powers, as shown in supplementary section Fig. A3.  
We have demonstrated that our sample exhibits macroscopic phase coherence \cite{shan2021coherent}, which rules out the purely excitonic behaviour. Considering the sample simultaneously presents the Zeeman splitting and macroscopic phase coherence, its polaritonic origin is unambiguously proved.
 
\subsection{Manifestation of the ground state brightening}\label{subsec3}

To demonstrate the brightening effect in the regime of the strong light-matter coupling, we compare the temperature-dependent PL for the isolated $\rm WSe_2$ flake and the flake placed inside the microcavity. The characteristic temperature-dependent exciton response has been employed previously to verify the conduction band inversion in $\rm MoWSe_2$ alloy monolayers \cite{wang2015spin}. 

We first turn our attention to the investigation of the bare exciton response of a pristine $\rm WSe_2$. To warrant a valid comparison, the $\rm WSe_2$ monolayer is exfoliated from the same crystal and transferred on an identical DBR as that used in our polariton sample. Again, the monolayer is capped by a thin h-BN layer. 
Throughout our experiments, we used a non-resonant continuous-wave laser at 532 nm focused to a spot of  $\sim$3 $\rm \mu m$ diameter to excite the sample (see more details of the setup in the Methods section). 

\begin{figure}[h]%
\centering
\includegraphics[width=0.9\textwidth]{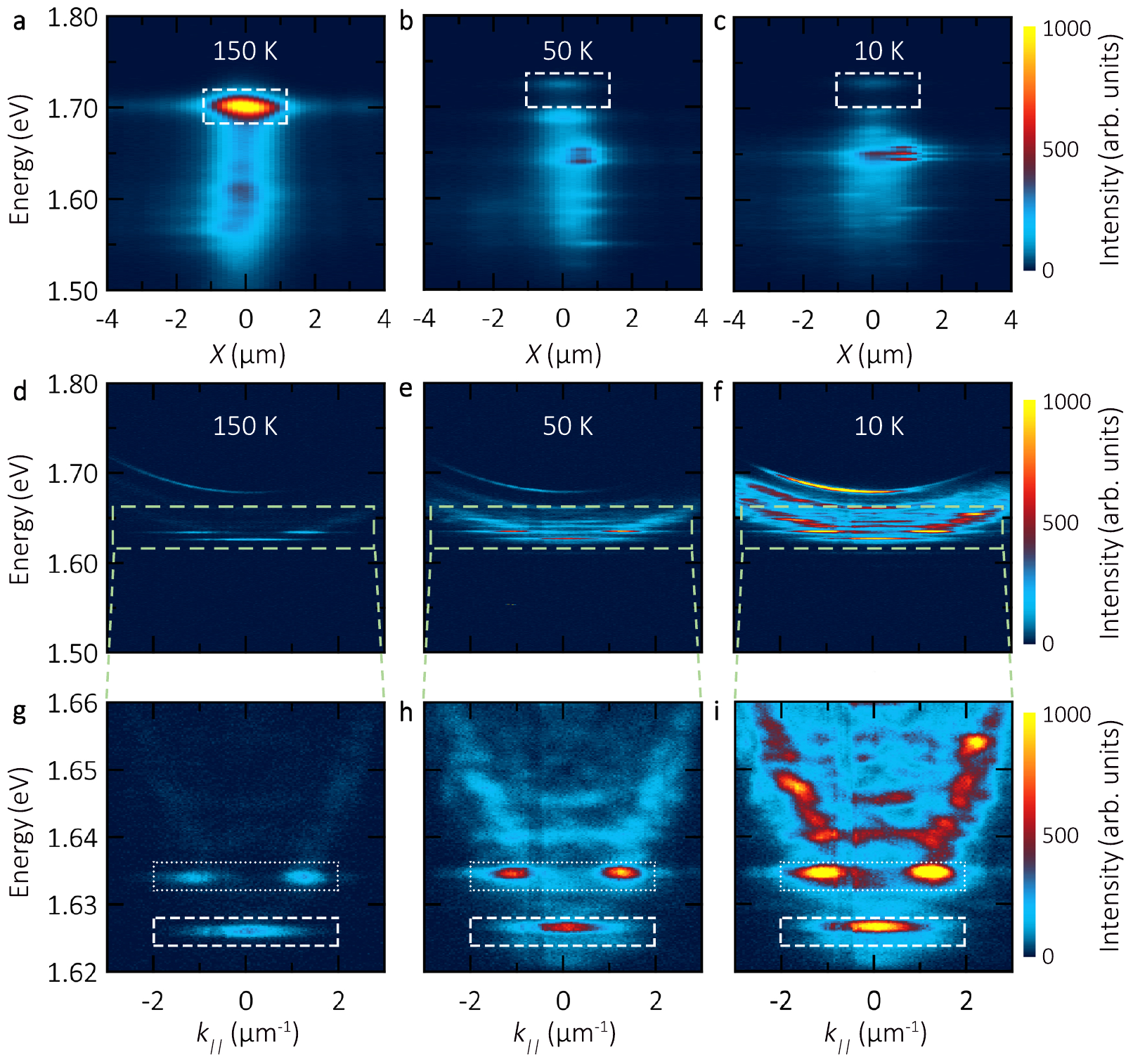}
\caption{\textbf {Temperature dependent PL of bare excitons and exciton-polaritons.} \textbf {a-c} Real-space resolved PL intensity distribution of a pristine $\rm WSe_2$ monolayer flake as a function of energy at temperatures 150, 50 and 10 K, respectively. The emission of bare excitons becomes dimmer as temperature decreases: the hallmark of a dark exciton ground-state. \textbf {d-f} Polariton dispersion relations recorded at 150, 50 and 10 K. In contrast to bare excitons, the LP luminescence significantly increases at low temperatures, behaving in the same manner as that of a bright material. \textbf {g-i} The zoom-in images of panels d-f in energy. The highlighted boxes are analysis regions of integrated intensity in Fig. 4.} \label{fig3}
\end{figure}

Figures 3a-c show the PL intensity distribution of the pristine $\rm WSe_2$ monolayer for three temperatures, 150, 50 and 10 K. The  pump power is kept at 30 $\rm \mu W$ (see more details in Methods section). At 150 K, one clearly sees strong PL signal from the flake region at the energy corresponding to the bright exciton with a broad emission tail at lower energies, which is probably associated with trions and spectrally broad defect-induced PL. As the temperature is reduced down to 50 K and finally 10 K, the exciton energy is blue-shifted by tens of meV \cite{yan2014photoluminescence}, and the corresponding emission intensity is reduced by a factor of 20 due to the fast non-radiative relaxation to the spin-forbidden dark ground state \cite{zhang2015experimental}. Note that also the dark exciton substantially blue-shifts upon temperature reduction.

The angle-resolved PL spectra for the $\rm WSe_2$ flake inside the microcavity at the same temperatures are presented in Figs. 3d-f. To show the polaritonic data clearly, the photon energy scale is zoomed-in in panels g-i. The LP energy position is mainly defined by the cavity mode, which is unaffected by the temperature changes, and thus the thermal blueshift of LP is severely reduced with respect to that of the bare exciton. The detailed analysis of the peak corresponding to the ground state is presented in supplementary section Fig. A4. Most interestingly, the LP photoluminescence strongly increases in intensity with decreasing temperatures: The maximal emission intensity of the polariton ground state at 10 K is 4 times larger than that at 150 K, while it is enhanced by 7 times for the first excited polariton state in the trap. This opposite temperature-dependence of PL clearly indicates that the ground state of our hybrid system becomes optically bright. 

In Fig. 4a, we present the integrated PL intensity from bare excitons in a pristine $\rm WSe_2$ flake as a function of temperature. The integrated areas are displayed as dashed boxes in Figs. 3a-c. As temperature is reduced from 200 K to 10 K, the PL intensity (green circles) experiences an exponential drop, stemming from the fast relaxation towards the dark ground state \cite{zhang2015experimental}. As shown in the left inset of Fig. 4a, the presence of dark excitons at energies that lie below bright excitons, suggests preferential population of the dark state with the reduction of temperature. The quenched light emission at low temperatures can be thermally activated following the Boltzman distribution \cite{zhang2015experimental}. The slight intensity reduction in the region from 200 K to 270 K was previously reported in flux-grown samples \cite{edelberg2019approaching}, and was attributed to phonon-induced non-radiative channels. The schematic is displayed as the right inset of Fig. 4a. With the increase of temperature, acoustic phonons start to participate in the scattering with bright excitons, yielding final states that are dark due to the momentum-forbidden condition. It is worth noting, that the precise trend is slightly dependent on the integration area, and there could be sample-to-sample differences, depending on the method of sample growth, exfoliation, etc, but the general phenomenon of a generic intensity reduction is universally present \cite{zhang2015experimental, edelberg2019approaching}.

\begin{figure}[h]%
\centering
\includegraphics[width=0.9\textwidth]{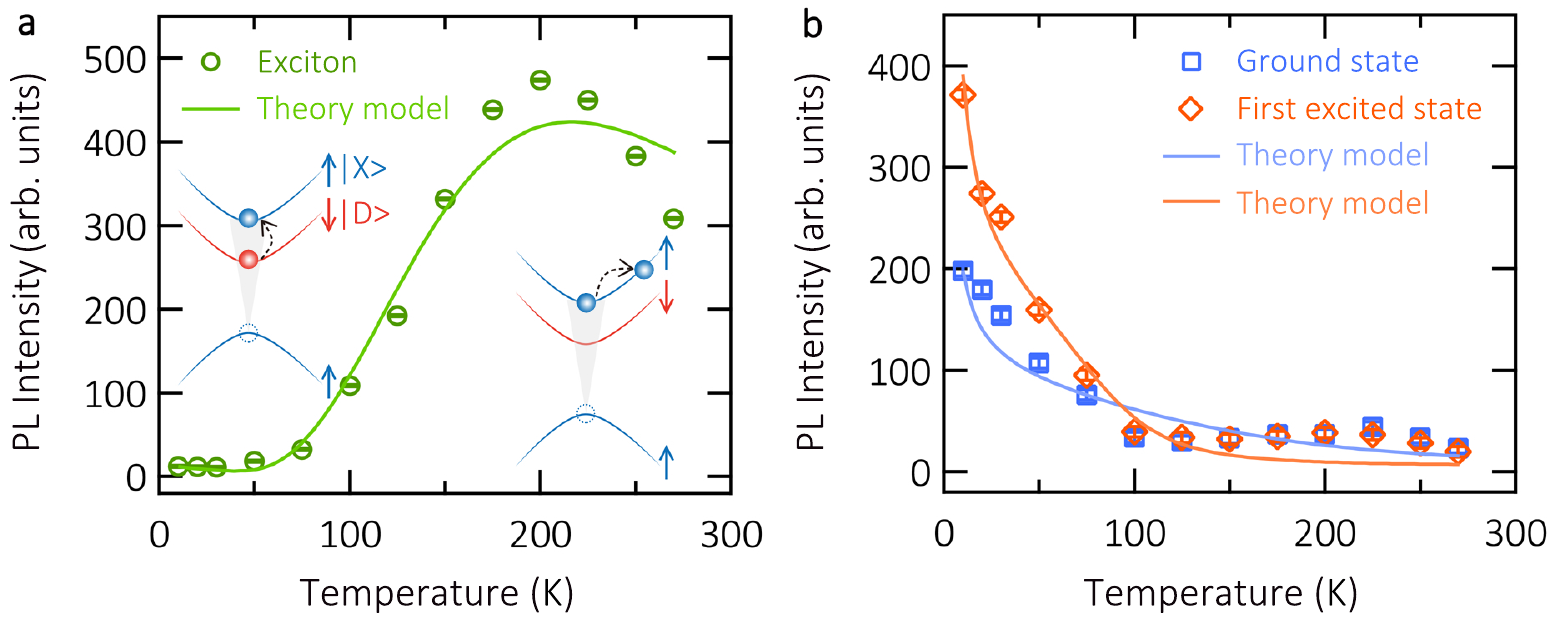}
\caption{\textbf {Integrated PL intensity of bare excitons and exciton-polaritons as a function of temperature.} \textbf {a} Temperature dependent emission intensity of bare excitons in pristine $\rm WSe_2$ monolayer. The experimental data are shown as green circles, and the corresponding integration regions are marked as dashed boxes in Figs. 3a-c. The solid curve represents the result of a theoretical modelling (see main text). The mechanism of PL intensity for different temperature ranges is shown as insets: thermal activation (10-200 K) and relaxation into momentum-forbidden dark states (200-270 K). \textbf {b} Temperature-dependent PL emission intensity of exciton-polaritons. The experimental data of the ground state and first excited state are shown as blue squares and red diamonds, respectively. The integration region of the ground state (first excited state) is indicated with a dashed (dotted) box in Figs. 3g-i. The solid curves are fits of the theory model. The strong PL intensity at low temperatures evidences the brightening effect of intrinsically dark exciton. The error bars are obtained by comparing the signal intensity to the standard deviation of the background noise.} \label{fig4}
\end{figure}

We modelled this temperature dependence of the bare exciton PL by solving the system of rate equations corresponding to the phonon-assisted population relaxation in a pristine $\rm WSe_2$ monolayer (see details in the supplementary section A.1). In this model, two kinetic equations are considered, which represent exciton occupancies of the bright and dark states. The bright exciton scatters into dark exciton by emitting a phonon, with a rate that depends on the bare exciton-phonon scattering rate $W_0$. The model fit is shown as a solid curve in Fig. 4a, closely reproducing the experimental trend. One clearly sees the thermal activation of the PL in the range from 10 to 200 K. Above this temperature, due to the relaxation into momentum-forbidden dark states that lie outside the light cone, the depletion of bright-exciton starts to play a role, which results in the slow decrease of PL in the range between 200 and 270 K.

The temperature dependence of the PL intensity in the strong coupling regime is presented in Fig. 4b, and clearly displays the opposite trend. The data for both trapped polariton states, ground state (blue squares) and the first excited state (red diamonds), are similarly analyzed and plotted. These polariton states show simultaneously a small intensity peak at around 225 K, which reproduces the behavior of bare excitons. Below 150 K the polariton PL dramatically enhances with reduced temperature, clearly demonstrating the brightening effect of the polariton ground state. Furthermore, we analyze the intensity of higher energy states that range from 1.64 to 1.66 eV in Fig. A6. Its temperature-dependence exhibits the same result as the ground and first excited states in Fig. 4b. Those higher energy states lie below the dark exciton, and polaritons relax into the ground state in a dynamic, cascaded manner. As such, it is natural to encounter increase of luminescence intensity of higher polariton states.

The results of the theoretical modelling based on a system of rate equations are shown by solid curves in Fig. 4b (see the details in the supplementary section A.2) and are in good quantitative agreement with the experimental data. In the simulations, we account for the ground and first excited cavity modes, and thus consider three polariton states. The resulting system contains four  equations describing the transitions between dark exciton  and  polariton branches (supplementary section A.2). The equation for the ground state occupancey reads:
\begin{align}
    &\dot{n}_L=-n_L/\tau_L+ X_L^2P_{eff}-W_{L\rightarrow D}(n_D+1)n_L+W_{D\rightarrow L}(n_L+1)n_D
\end{align}
where $\tau_{L}^{-1}=X_{L}^2\tau_{nr}^{-1}+C1_{L}^2\tau_{c1}^{-1}+C2_{L}^2\tau_{c2}^{-1}$,  $X_L,C1_L,C2_L$ are the corresponding Hopfield coefficients, and $W$ are the phonon-assisted inelastic scattering rates, the expressions for which are presented in the supplementary section A.2.

To account for the fast thermalization of non-resonantly pumped excitons, which leads to the decrease of the fraction of the thermal excitons lying within the light cone and forming polaritons with temperature, we introduce the effective pumping $P_{eff}$ which is expressed as:
\begin{align}
P_{eff}\approx P\frac{\epsilon_{lc}}{k_BT},
\end{align}
where $\epsilon_{lc}$ is the energy of the excitons corresponding to the wavevector of light in the material,
\begin{align}
    \epsilon_{lc}=\frac{\hbar^2}{2M_{exc}}\varepsilon_{TiO_2}(\omega_x/c)^2\approx 58~\mu eV\approx 0.6K.
\end{align}
This approximation holds for $T\gg 0.6$K. 

As lower polariton states at around $ k_{\parallel}=0 $ have energies far below the dark exciton energy, LP can be effectively populated by inelastic processes of phonon emission. Thus, differently from the case of bare excitons, most of the bright excitons contribute to the PL signal via polariton emission at low temperatures, which explains the rapid increase of PL intensity with the decreasing temperature.
In this work, we study temperature dependent polariton emission while the Rabi-splitting remains approximately constant. Similarly, we believe that tuning of the Rabi-splitting, e.g. using an open-cavity \cite{flatten2016room} at constant temperature could be an alternative method to probe the brightening effect.

In strongly coupled organic microcavities, the reverse intersystem crossing rates are unchanged because of the delocalization nature of polaritons \cite{eizner2019inverting, martinez2019triplet}. In this work, however, we find the strong coupling is a feasible approach. Two possible reasons may be attributed: (1) the spatial confinement of polaritons, which facilitates polaritonic relaxation to the ground state via phonons, (2) the high quality factor of optical cavity, which prolongs lifetimes of polaritons.

\section{Discussion}\label{sec3}

In summary, we demonstrate the possibility to effectively brighten an intrinsically dark semiconductor monolayer by placing it inside a resonantly tuned optical
microcavity in the regime of strong light-matter coupling. 

In our experiments, we utilize a $\rm WSe_2$ monolayer flake, which features the spin-forbidden dark exciton ground state, separated from the bright state by an energy of $\sim 40$ meV. This splitting is comparable with the vacuum Rabi-splitting characteristic to the resonant coupling of bright excitons with cavity photons, which allows to push the energy of the lower polariton below the energy of the dark exciton. As a consequence, the ground state of the system becomes bright, and instead of showing a PL quenching, characteristic for pristine $\rm WSe_2$ monolayers, we observe a strong PL enhancement with decreasing temperature, due to the fast and efficient phonon-assisted energy relaxation.

 In a broader scale, our approach for band-structure engineering aligns with contemporary efforts to tune transport, topological and magnetic properties of low dimensional materials using the fundaments of cavity quantum electrodynamics. 
 
\section{Methods}\label{sec4}

\textbf{Sample fabrication.} The bottom DBR, composed of 10 pairs of $\rm SiO_2/TiO_2$ layers, is obtained commercially from Laseroptics GmbH. The mechanically exfoliated $\rm WSe_2$ monolayer is capped by a $\sim$5 $\rm nm$ h-BN multilayer via the deterministic dry-transfer method. The top DBR is evaporated by an ion-assisted physical vapor deposition process \cite{knopf2019integration}, consisting of 9 pairs of $\rm SiO_2/TiO_2$ layers: the $\rm SiO_2$ layer in contact with h-BN is 105 nm in thickness, and the thickness of each repetitive pair is 129 and 83 nm, respectively (more details are described in Ref. \cite{shan2021coherent}).
The control sample constituted by a pristine $\rm WSe_2$ monolayer is also mechanically exfoliated from a bulk crystal. It is dry-transferred onto another identical DBR. An h-BN multilayer with similar thickness ends the final capping.
\newline
\textbf{Experimental setup.} A standard back Fourier plane imaging setup is utilized to perform angle-resolved PL measurements. The excitation source is a continuous-wave (CW) 532 nm laser, and it is focused on the sample by a long working-distance objective (Mitutoyo M Plan Apo NIR 50x, NA=0.42). The charge-coupled device (CCD) of Andor (iDus 416) is attached to a spectrometer (Shamrock 500i). Its sensor is operated at -80 $\rm ^oC$. A 600 nm longpass filter of Thorlabs (FELH0600) is used to block the green laser from reaching the CCD. In the measurements of Figs. 3a-c, the pump power is kept at 30 $\rm \mu W$, and the exposure time is 2 s. In the measurements of Figs. 3d-f, the pump power is 30 $\rm \mu W$ and the exposure time is increased to 10 s, to compensate for the less efficient light-collection in the back-fourier plane imaging configuration. To carry out temperature-dependent experiments, the sample is mounted on a motorized XY stage of a customized low-vibration cryostat from ColdEdge. The available lowest temperature is 10 K, and the temperature of the sample is adjusted by a Lakeshore temperature controller (Model 335). The magnetic measurements are performed at room temperature using an attoDRY2100. A CW 532 nm laser excites the sample, and a set of quarter waveplate, half waveplate as well as linear polarizer in the collection path is used to detect circularly-polarized PL emission. In magnetic field measurements, the used pump power is 100 $\rm \mu W$ and the exposure time is 120 s.

\bmhead{Data availability}
The data that support the findings of this study are available from the corresponding authors upon reasonable request.

\bibliography{sn-bibliography}% common bib file


\backmatter

\section*{Declarations}
\bmhead{Supplementary information}
The supplementary file of article is attached as Appendix A. 

\bmhead{Acknowledgments}
The authors gratefully acknowledge funding by the State of Lower Saxony. Funding provided by the European Research Council (ERC project 679288, unlimit-2D) is acknowledged. I.I., S.H. and I.A.S. acknowledge the support from the joint RFBR-DFG project No. 21-52-12038. I.I. acknowledges the support Ministry of Science and Higher Education of Russian Federation, goszadanie no. 2019-1246. S.T. acknowledges funding from NSF DMR 1955889, DMR 1933214, and 1904716. S.T. also acknowledges DOE-SC0020653, DMR 2111812, and ECCS 2052527 for material development and integration. K.W. and T.T. acknowledge support from the Elemental Strategy Initiative conducted by the MEXT, Japan (Grant Number JPMXP0112101001) and JSPS KAKENHI (Grant Numbers JP19H05790 and JP20H00354). H.S. acknowledges the Sino-Germany (CSC-DAAD) Postdoctoral Scholarship Program from China Scholarship Council and German Academic Exchange Service. S.K. and S.H. acknowledge the Deutsche Forschungsgemeinschaft (DFG, German Research Foundation)–INST 93/932-1 FUGG.  S.H. acknowledges financial support by The German Research Foundation (DFG) (HO 5194/16-1). M.E. acknowledges funding by the University of Oldenburg through a Carl-von-Ossietzky fellowship. F.E. and H.K. are supported by the Federal Ministry of Education and Science of Germany under Grant ID 13XP5053A.

\bmhead{Author contributions}
This work was initialized and guided by C.S. and C.A.-S. The experiments were conducted by H.S., B.H. and C.A.-S. The magnetic field measurements were carried out by H.S. and C.A.-S., supervised by S.K. and S.H. The data analysis was performed by H.S., C.S., M.E. and C.A.-S. The TMDC [h-BN] crystals were produced by K.Y. and S.T. [K.W. and T.T.]. The sample was fabricated by C.R., and further processed by H.K. and F.E. The theory and simulations were performed by I.I., under the supervision of I.A.S. The manuscript was written by C.S., H.S.,  I.I., I.A.S. and C.A.-S., with input from all the co-authors.

\bmhead{Competing  interests}
The authors declare no competing interests.

\bmhead{Figure legends}

\textbf {Fig. 1 Sample structure and brightening mechanism.} \textbf {a} Schematic illustration of sample structure with a $\rm WSe_2$ monolayer, capped with h-BN, and embedded between two dielectric DBRs. Two arrows represent the Rabi oscillation between excitons and microcavity photons. \textbf {b} Optical microscopic image of the sample. The $\rm WSe_2$ monolayer and h-BN boundaries are indicated with yellow and white dashed lines, respectively. \textbf {c} Scheme of $\rm WSe_2$ ground state brightening via strong coupling. In pristine $\rm WSe_2$ monolayers, the optically dark exciton $\rm \lvert D \rangle $ (brown line) is the lowest transition at the $K$-point, which lies $\sim$40 meV below the bright exciton $\rm \lvert X \rangle $ (orange line). Exciton-polaritons are formed when optically bright excitons strongly couple to microcavity photons, the corresponding energy diagram is enclosed in a dashed box. The energy level of the resulting lower polaritons $\rm \lvert LP \rangle $ can be located below the dark exciton state  $\rm \lvert D \rangle $, as long as the Rabi splitting $\Omega_R$ is sufficiently large. The lower polariton branch, which inherits the spin character from the bright exciton, becomes the ground state of the coherently dressed system. Thus, the band ordering is reversed, and the intrinsically dark 2D semiconductor is effectively brightened via the strong coupling with microcavity photons.
\\
\textbf {Fig. 2 Polariton dispersion relation and valley-Zeeman effect.} \textbf {a} Dispersion relation of exciton-polaritons at ambient conditions. The exciton (X) and microcavity photon (C) are represented by dotted and dashed lines, respectively. The dark exciton is marked as D. The upper and lower polariton (UP and LP) branches are plotted as solid lines. The discrete energy modes of LP are a typical dispersion relation of polaritons confined at a finite-size trap. \textbf {b} Normalized circularly-polarized PL intensity spectra of  exciton-polaritons under a magnetic field at room temperature. These spectra are extracted from the dispersion relations at zero in-plane momentum $k_{\parallel}{=}0$. From top to bottom panels, the applied magnetic field is +9, 0 and -9 T, respectively. $\rm \sigma^+$ ($\rm \sigma^-$) denotes emission of light with right (left)-hand circular polarization. An energy splitting is observed under the application of magnetic fields.
\\
\textbf {Fig. 3 Temperature dependent PL of bare excitons and exciton-polaritons.} \textbf {a-c} Real-space resolved PL intensity distribution of a pristine $\rm WSe_2$ monolayer flake as a function of energy at temperatures 150, 50 and 10 K, respectively. The emission of bare excitons becomes dimmer as temperature decreases: the hallmark of a dark exciton ground-state. \textbf {d-f} Polariton dispersion relations recorded at 150, 50 and 10 K. In contrast to bare excitons, the LP luminescence significantly increases at low temperatures, behaving in the same manner as that of a bright material. \textbf {g-i} The zoom-in images of panels d-f in energy. The highlighted boxes are analysis regions of integrated intensity in Fig. 4.
\\
\textbf {Fig. 4 Integrated PL intensity of bare excitons and exciton-polaritons as a function of temperature.} \textbf {a} Temperature dependent emission intensity of bare excitons in pristine $\rm WSe_2$ monolayer. The experimental data are shown as green circles, and the corresponding integration regions are marked as dashed boxes in Figs. 3a-c. The solid curve represents the result of a theoretical modelling (see main text). The mechanism of PL intensity for different temperature ranges is shown as insets: thermal activation (10-200 K) and relaxation into momentum-forbidden dark states (200-270 K). \textbf {b} Temperature-dependent PL emission intensity of exciton-polaritons. The experimental data of the ground state and first excited state are shown as blue squares and red diamonds, respectively. The integration region of the ground state (first excited state) is indicated with a dashed (dotted) box in Figs. 3g-i. The solid curves are fits of the theory model. The strong PL intensity at low temperatures evidences the brightening effect of intrinsically dark exciton. The error bars are obtained by comparing the signal intensity to the standard deviation of the background noise.
\bigskip

\newpage

\begin{appendices}

\pagestyle{empty}

\section{Supplementary information}\label{secA1}
\subsection{Model of temperature-dependent exciton photoluminescence of $\rm WSe_2$ monolayer}

To model the temperature dependent PL signal of the uncoupled exciton in WSe$_2$ monolayer, we solve the system of two kinetic equations for the exciton occupancies $n_B,n_D$ in the bright and dark states, respectively:
\begin{align}
    &\dot{n}_B=P_{eff}-n_B/\tau_R-n_B/\tau_{NR}-W_{B\rightarrow D} n_B(1+n_D)+W_{D\rightarrow B} n_D(1+n_B),\\
    &\dot{n}_D=-n_D/\tau_{NR}+W_{B\rightarrow D} n_B(1+n_D)-W_{D\rightarrow B} n_D(1+n_B),
\end{align}
where $P_{eff}$ defines the effective incoherent pumping of the bright excitons. We assume that as pumped incoherently, the excitons thermalize very quickly, and only the excitons lying within the light cone contribute to the PL signal. For not very low temperature, the effective pumping $P_{eff}$ is
\begin{align}
P_{eff}\approx P\frac{\epsilon_{lc}}{k_BT},
\end{align}
where 
\begin{align}
    \epsilon_{lc}=\frac{\hbar^2}{2M_{exc}}\varepsilon_{TiO_2}(\omega_x/c)^2\approx 58~\mu eV\approx 0.6K.
\end{align}
This approximation holds for $T\gg 0.6$K. Scattering rates $W$ correspond to the inelastic phonon-assisted process. Since dark excitonic states lie below the bright ones, the bright to dark exciton scattering occurs with the emission of a phonon, and a reverse process with an absorption of the phonon. Assuming the system is at thermal equilibrium at temperature T, the scattering rates can be expressed as:
\begin{align}
    &W_{B\rightarrow D} = \frac{W_0}{1-\exp[-\Delta_{BD}/k_BT]},\\
    &W_{D\rightarrow B}= W_{B\rightarrow D} e^{-\Delta_{BD}/k_BT},
\end{align}
where $\Delta_{BD}$ is the bright-dark exciton splitting which is equal to the spin splitting in the conduction band and is equal to approximately 40 meV~\cite{zhou2017probing, wang2017in-plane}, and $W_0$ is the bare exciton-phonon scattering rate which is a fitting parameter. The radiative exciton lifetime of the excitons within the light cone is set to $4$ ps which corresponds to previously reported theoretical and experimental values~\cite{wang2014valley,palummo2015exciton} and the non-radiative lifetime which is typically related to the defect trapping is set to  $\tau_{NR}=5$~ps~\cite{korn2011low}. By fitting the experimental temperature dependence we obtain the value of $W_0\approx 10~\mathrm{ps}^{-1}$.

\subsection{Model of ground state brightening via strong light-matter coupling}

We consider a simplified picture, where there is a single exciton state, and two cavity modes, corresponding to the ground and excited exciton modes in the trap (Fig. A1). We can write down the coherent part of the Hamiltonian in the matrix form
\begin{align}
    H_{coh}= \begin{pmatrix} b^{\dagger} & a_1^{\dagger} & a_2^{\dagger}  \end{pmatrix} \begin{pmatrix} \omega_x & g_1 & g_2 \\ g_1 & \omega_{c1} & 0 \\ g_2 & 0 & \omega_{c2}\end{pmatrix}\begin{pmatrix} b \\ a_1 \\ a_2\end{pmatrix}
\end{align}
where $b,a_1,a_2$ are the annihilation operators for the exciton and two cavity modes. The frequencies $\omega_x$, $\omega_{c1},\omega_{c2}$ can be extracted from the experimental data, and $g_1,g_2$ are the fitting parameters in our model.

\begin{figure}[h]%
\centering
\includegraphics[width=0.7\textwidth]{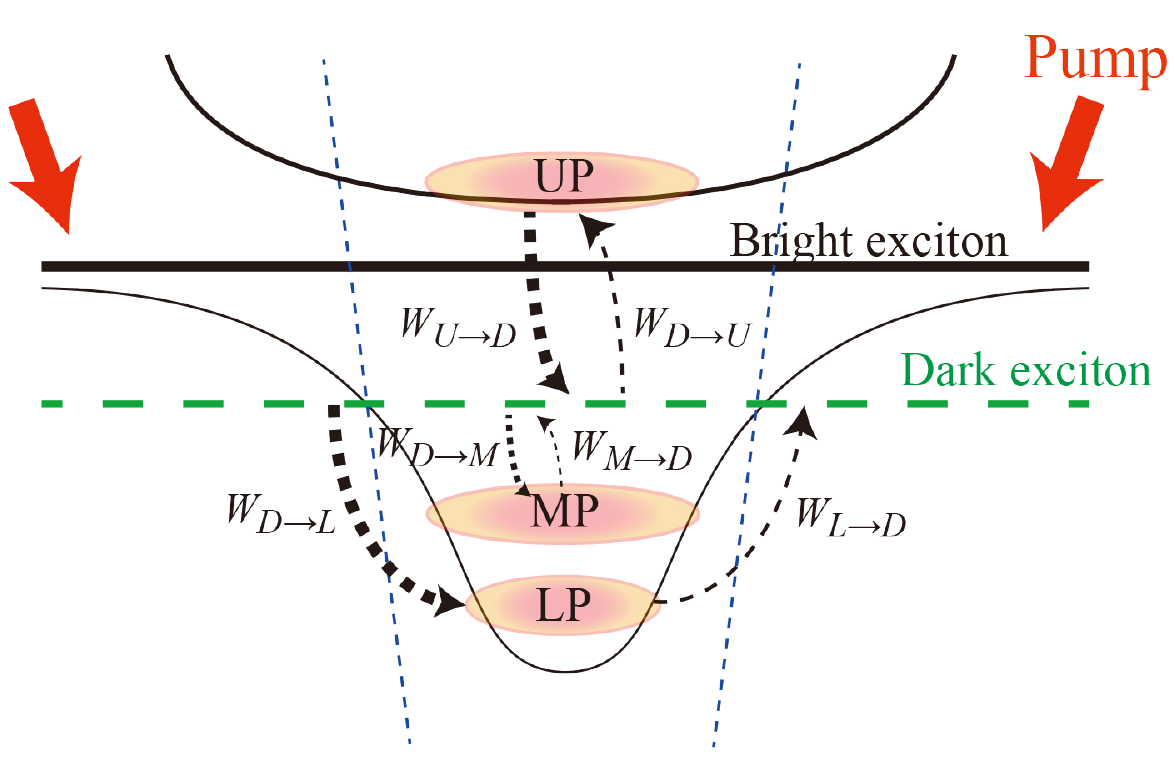}
\caption{\textbf {Scheme of phonon-assisted population relaxation of the hybrid system.}  In the simulations, we consider four population equations for the dark excitons as well as three polariton branches LP, MP and UP. The expressions of scattering rates of corresponding transitions are written as Equations. (A8-13). }\label{fig S1}
\end{figure}

This Hamiltonian can be diagonalized yielding lower, middle and upper polariton modes $a_{L},a_{M}, a_{U}$. Moreover, we obtain the polariton energies $\epsilon_L,\epsilon_M,\epsilon_U$ and exciton fraction in each of the polaritonic modes (Hopfield coefficients): $X_U,X_M,X_L$.
The rate of the phonon assisted scattering between the polaritons and the spin-dark excitons $D$ is proportional to the exciton fraction in the respective polariton, and the occupation number of the phonons at the energy corresponding to the energy difference between the dark exciton and respective polariton. If the scattering occurs with the emission of phonon, the rate is proportional to $n_{ph}+1$ and if the scattering is with the absorption of phonon, then the rate is proportional to $n_{ph}$. 
The phonon scattering rates thus are:
\begin{align}
    &W_{U\rightarrow D}=X_U^2 W_0\frac{1}{1-\exp[-(\epsilon_U-\epsilon_D)/(k_BT)]},\\
    &W_{D\rightarrow U}=e^{-(\epsilon_U-\epsilon_D)/(k_BT)}W_{U\rightarrow D},\\
    &W_{D\rightarrow M}=X_M^2 W_0\frac{1}{1-\exp[-(\epsilon_D-\epsilon_U)/(k_BT)]},\\
    &W_{M\rightarrow D}=e^{-(\epsilon_D-\epsilon_U)/(k_BT)}W_{D\rightarrow M},\\
    &W_{D\rightarrow L}=X_L^2 W_0\frac{1}{1-\exp[-(\epsilon_D-\epsilon_L)/(k_BT)]},\\
    &W_{L\rightarrow D}=e^{-(\epsilon_D-\epsilon_L)/(k_BT)}W_{D\rightarrow L},
\end{align}
where $W_0$ is the bare rate of exciton-phonon scattering obtained via fitting of the bare exciton PL signal. 
We also need to account for the fact that at non-resonant pumping only the excitons with the wavevectors inside the light cone form the polaritons. 
The system of equation then reads
\begin{align}
    &\dot{n}_U=-n_U/\tau_U + X_U^2P_{eff}-W_{U\rightarrow D}(n_D+1)(n_U)+W_{D\rightarrow U}n_D(n_U+1),\\
    &\dot{n}_M=-n_M/\tau_M+ X_M^2P_{eff}-W_{M\rightarrow D}(n_D+1)n_M+W_{D\rightarrow M}(n_M+1)n_D,\\
    &\dot{n}_L=-n_L/\tau_L+ X_L^2P_{eff}-W_{L\rightarrow D}(n_D+1)n_L+W_{D\rightarrow L}(n_L+1)n_D,\\
    &\dot{n}_D=-n_D/\tau_D+ W_{U\rightarrow d}(n_D+1)(n_U)-W_{D\rightarrow U}n_D(n_U+1)+W_{M\rightarrow D}(n_D+1)n_M-\nonumber\\&-W_{D\rightarrow M}(n_M+1)n_D+W_{L\rightarrow D}(n_D+1)n_L-W_{D\rightarrow L}(n_L+1)n_D
\end{align}
where $\tau_{U,M,L}^{-1}=X_{U,M,L}^2\tau_x^{-1}+C1_{U,M,L}^2\tau_{c1}^{-1}+C2_{U,M,L}^2\tau_{c2}^{-1}$, and $\tau_D$ is the dark exciton lifetime which is set equal to the non-radiative exciton lifetime. The values of $g_1=0.02~\mathrm{eV}$ and $g_2=0.016~\mathrm{eV}$ are extracted from fitting the experimental temperature dependence of the polariton PL signal. Moreover, in fitting we accounted for the temperature dependence of the bare exciton energy $\omega_x\approx 1.73-0.06(T/300)$ eV.

\subsection{Polariton dispersion relations versus air pressure}

\begin{figure}[H]%
\centering
\includegraphics[width=0.9\textwidth]{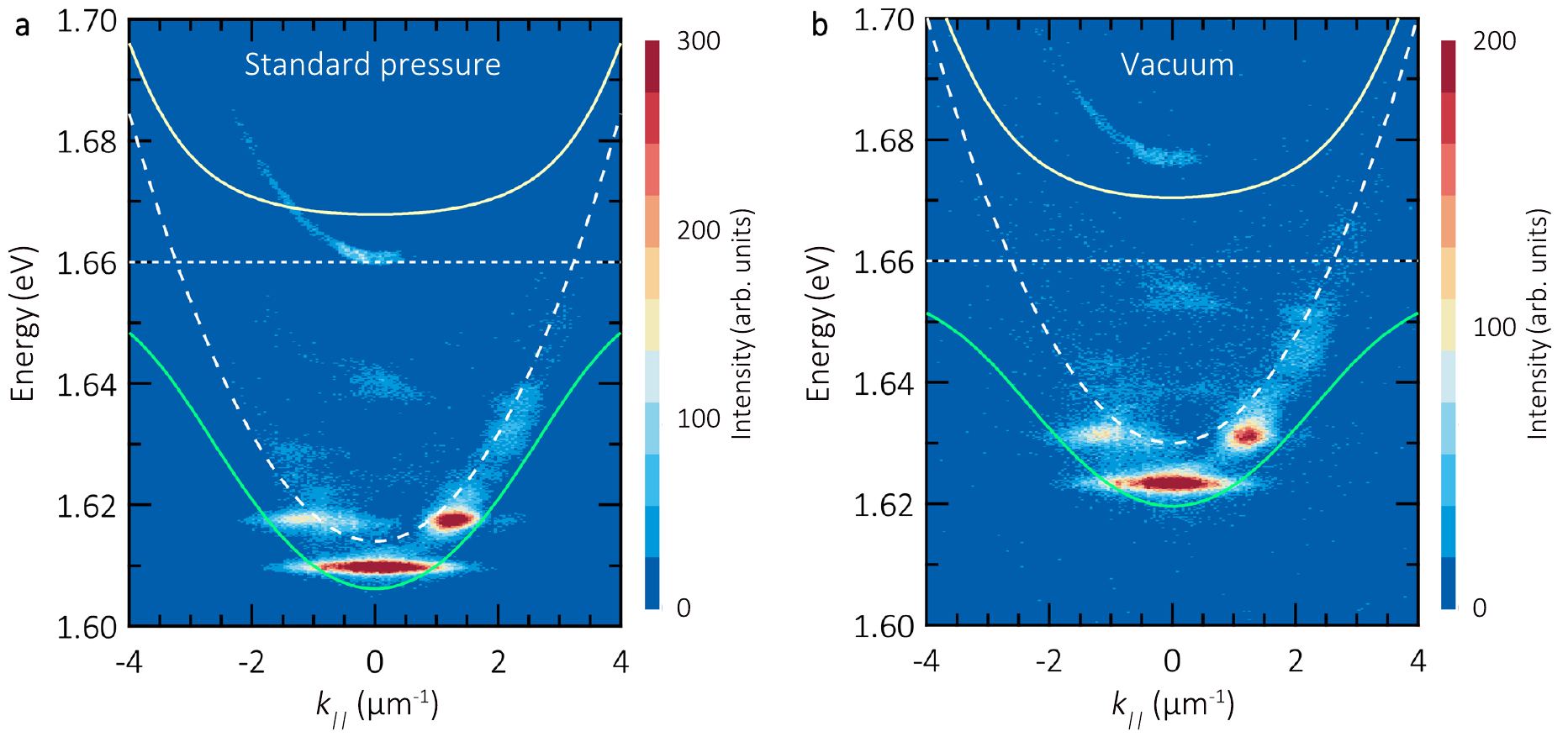}
\caption{\textbf {Dispersion relations of room temperature polaritons at standard pressure and high vacuum.} \textbf {a} Polariton dispersion relation at ambient conditions, which is reproduced from Fig. 2a. The bare cavity mode is at 1.614 eV. \textbf {b} Polariton dispersion relation at room temperature and high vacuum. The cavity mode is changed to 1.630 eV as the pressure is decreased to 3E-5 hPa.}\label{fig S2}
\end{figure}

Air pressure can affect the energy of microcavity photons and LP. Figure A2a shows the polariton dispersion at ambient conditions (room temperature, standard pressure), which is reproduced from Fig. 2a. The cavity mode is at 1.614 eV, and the energy of ground state is 1.610 eV.

Weak emission detected above 1.66 eV is collected from areas without the monolayer-hBN stack, i.e. the barrier of the polariton trap (the collection area of PL is tens of microns). Since the optical length of the cavity is significantly reduced at those positions, it occurs at strongly elevated energy.In the barrier, there is no active material which can directly generate luminescence, however light, which is generated by the polaritons in the trap or which is generated by weakly coupled excitons on the lateral interface between trap and barrier, can scatter into these modes.

Before performing temperature-dependent measurements, the cryostat is pumped several hours to get a high vacuum. Figure A2b shows the dispersion relation after vacuum pump (room temperature, pressure of 3E-5 hPa). We observe an energy shift of LP, which originates from the alteration of cavity mode. The cavity photon energy is changed to 1.630 eV and the resulted  LP is at 1.624 eV. This energy shift is reversible: the dispersion relation gets back after measurements, when the air pressure is recovered.

\subsection{The valley-Zeeman effect at low pump power}

\begin{figure}[H]%
\centering
\includegraphics[width=0.9\textwidth]{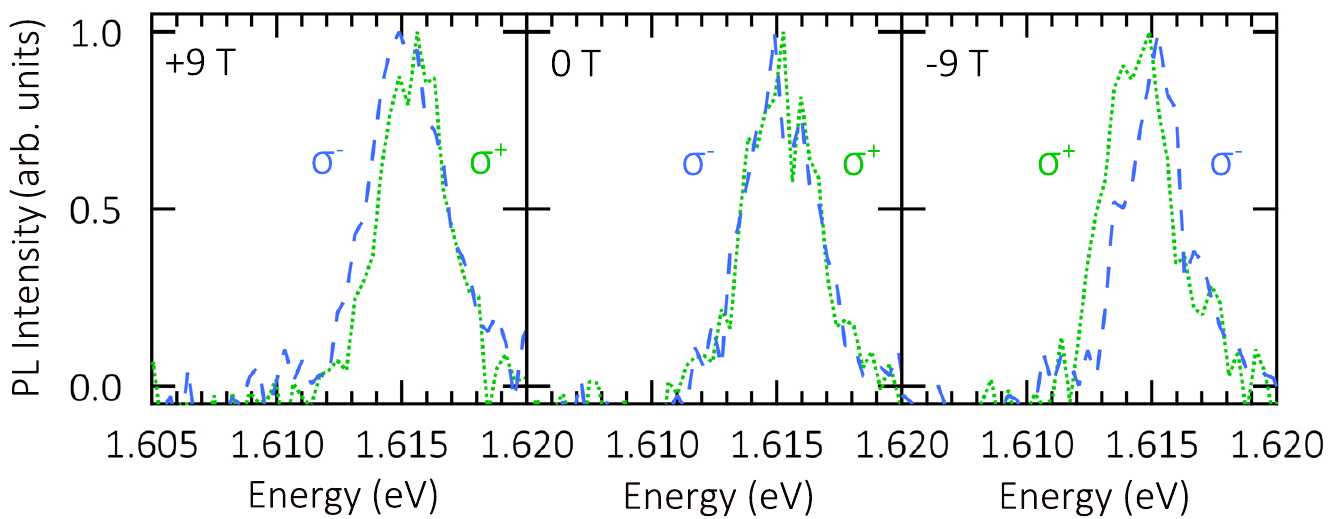}
\caption{\textbf {Normalized circularly-polarized PL intensity of  exciton-polaritons under a magnetic field at room temperature.} The pump power is 10 $\rm \mu W$. From left to right panels, the applied magnetic field is +9, 0 and -9 T. $\rm \sigma^+$ ($\rm \sigma^-$) denotes emission light of right (left)-hand circular polarization.}\label{fig S3}
\end{figure}

\subsection{Supplied temperature dependent data of exciton-polaritons}

\begin{figure}[h]%
\centering
\includegraphics[width=0.5\textwidth]{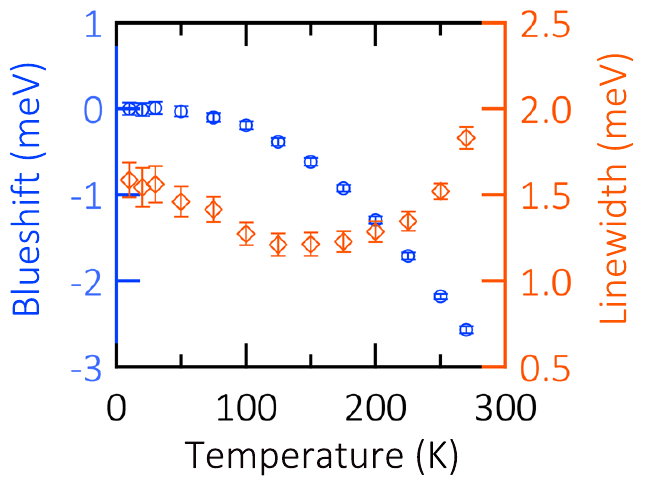}
\caption{\textbf {Blueshift (blue circles) and linewidth (red diamonds) of polariton ground state as a function of temperature.} The energy of polariton ground state is blueshifted $\sim$3 $\rm meV$ cooling from 270 K to 10 K. The error bars of blueshift and linewidth correspond to the 95\% confidence interval of the peak fitting.}\label{fig S4}
\end{figure}

\begin{figure}[H]%
\centering
\includegraphics[width=0.9\textwidth]{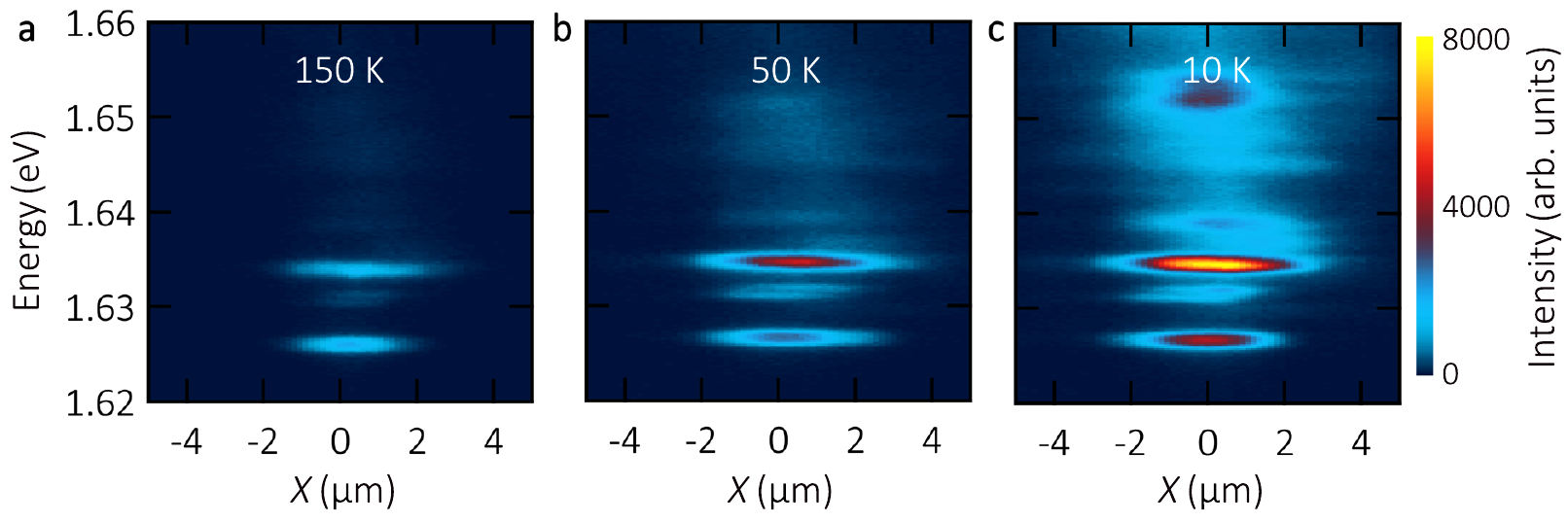}
\caption{\textbf {Real-space resolved PL intensity distribution of polaritons as a function of energy.} \textbf {a-c} The temperature is 150 K, 50 K and 10 K. The emission intensity of polaritons in real-space agrees with the trend of momentum-space as shown in Fig. 3d-f.}\label{fig S5}
\end{figure}

\begin{figure}[h]%
\centering
\includegraphics[width=0.8\textwidth]{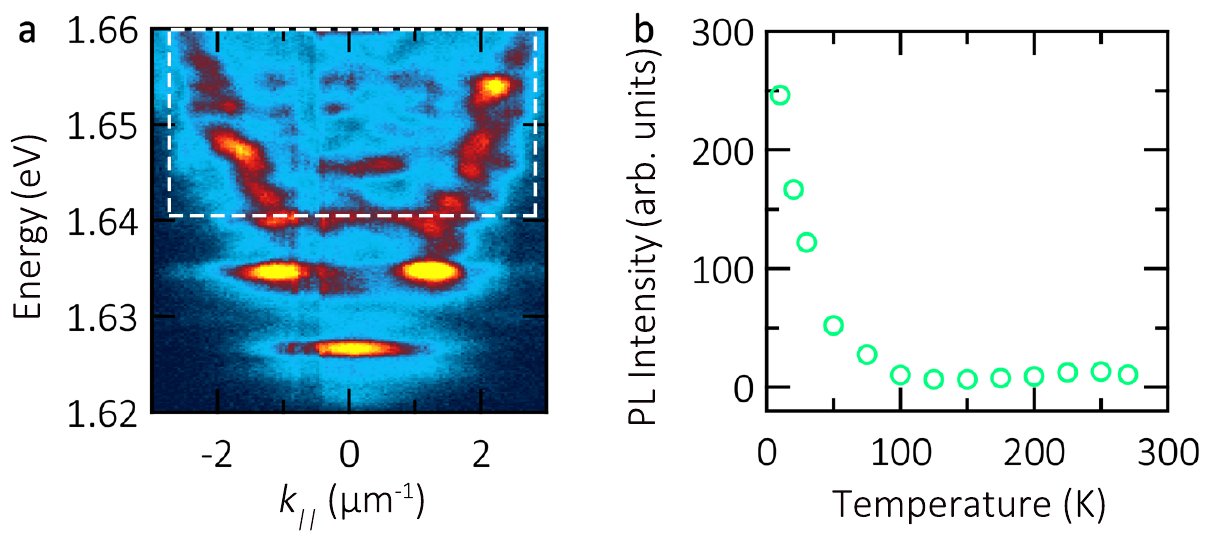}
\caption{\textbf {Integrated PL intensity of polaritons.} \textbf {a} is reproduced from Fig. 3i. The white box represents the integration region that ranges from 1.64 to 1.66 eV. \textbf {b} Temperature dependent emission intensity of polaritons, with the integration region shown in a.}\label{fig S6}
\end{figure}

\begin{figure}[h]%
\centering
\includegraphics[width=0.8\textwidth]{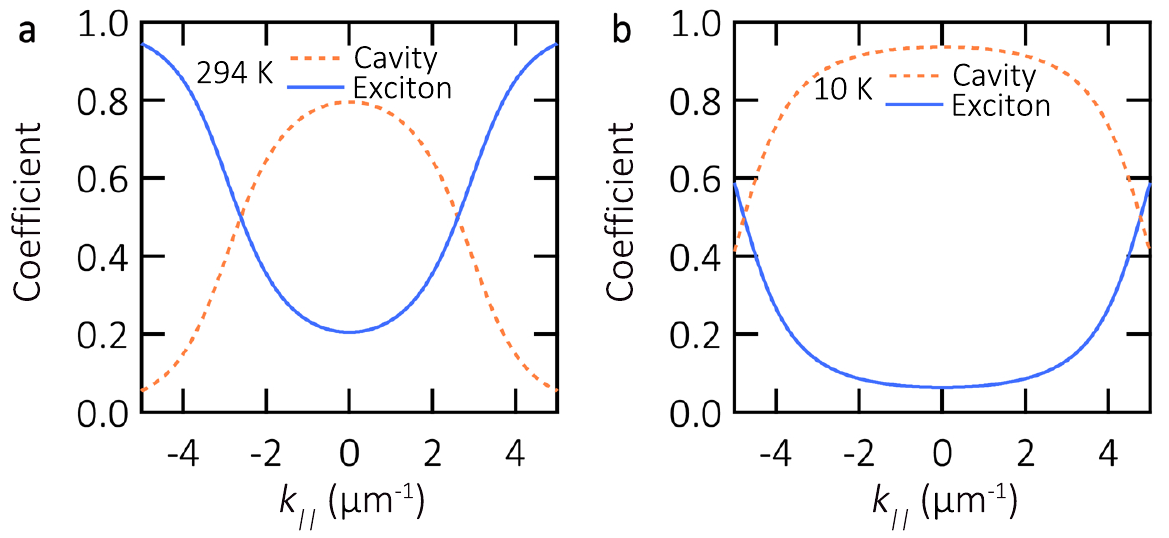}
\caption{\textbf {Hopfield coefficients of LP at temperatures of 294 K (a) and 10 K (b).} The exciton fraction of LP is 20.5\% at 294 K, and it decreases to 6.4\% as the temperature is 10 K.}\label{fig S7}
\end{figure}

\subsection{Supplied temperature dependent data of pristine $\rm WSe_2$ monolayer}

\begin{figure}[h]%
\centering
\includegraphics[width=0.9\textwidth]{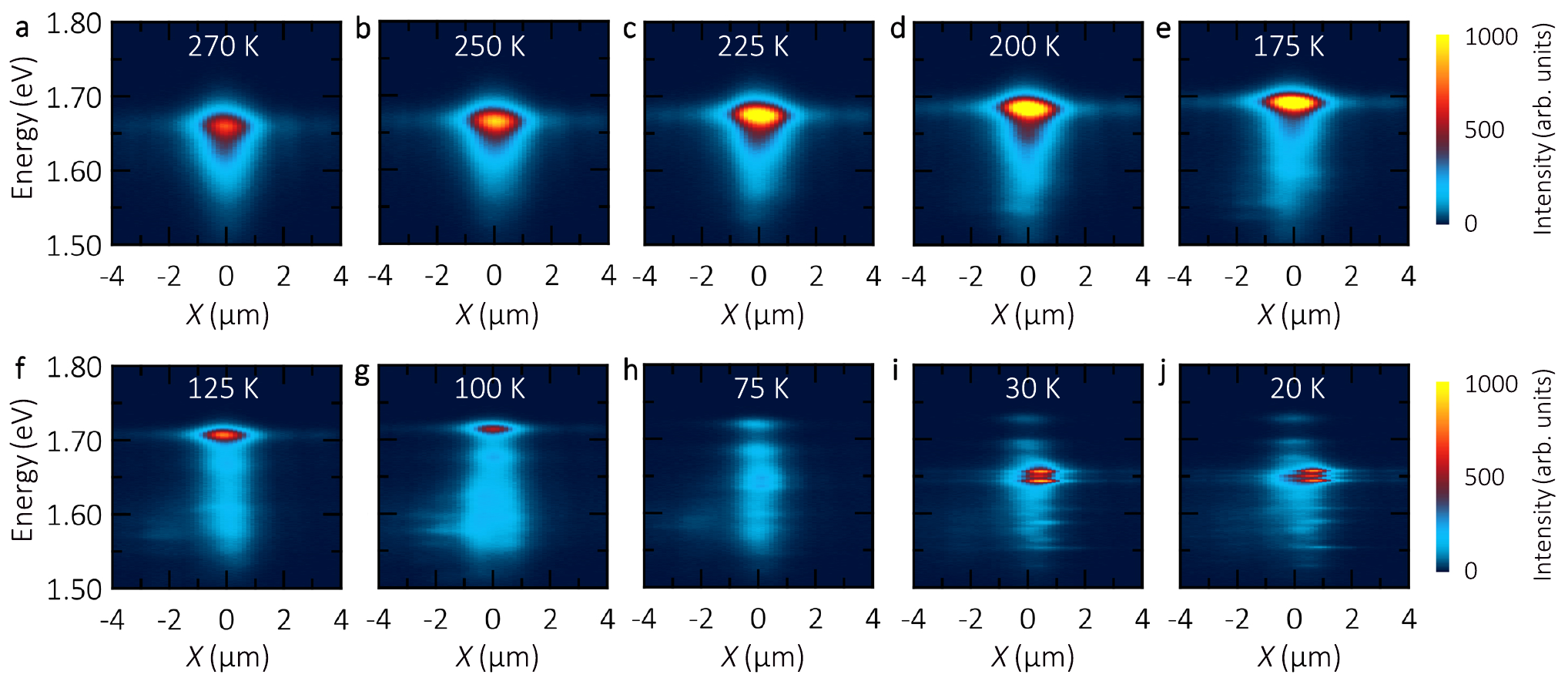}
\caption{\textbf {Real-space resolved PL intensity distribution of pristine $\rm WSe_2$ monolayer as a function of energy.} The temperature decreases from 270 K to 20 K. The rest data of 150 K, 50 K and 10 K are shown in Fig. 3a-c.}\label{fig S8}
\end{figure}

\begin{figure}[h]%
\centering
\includegraphics[width=0.6\textwidth]{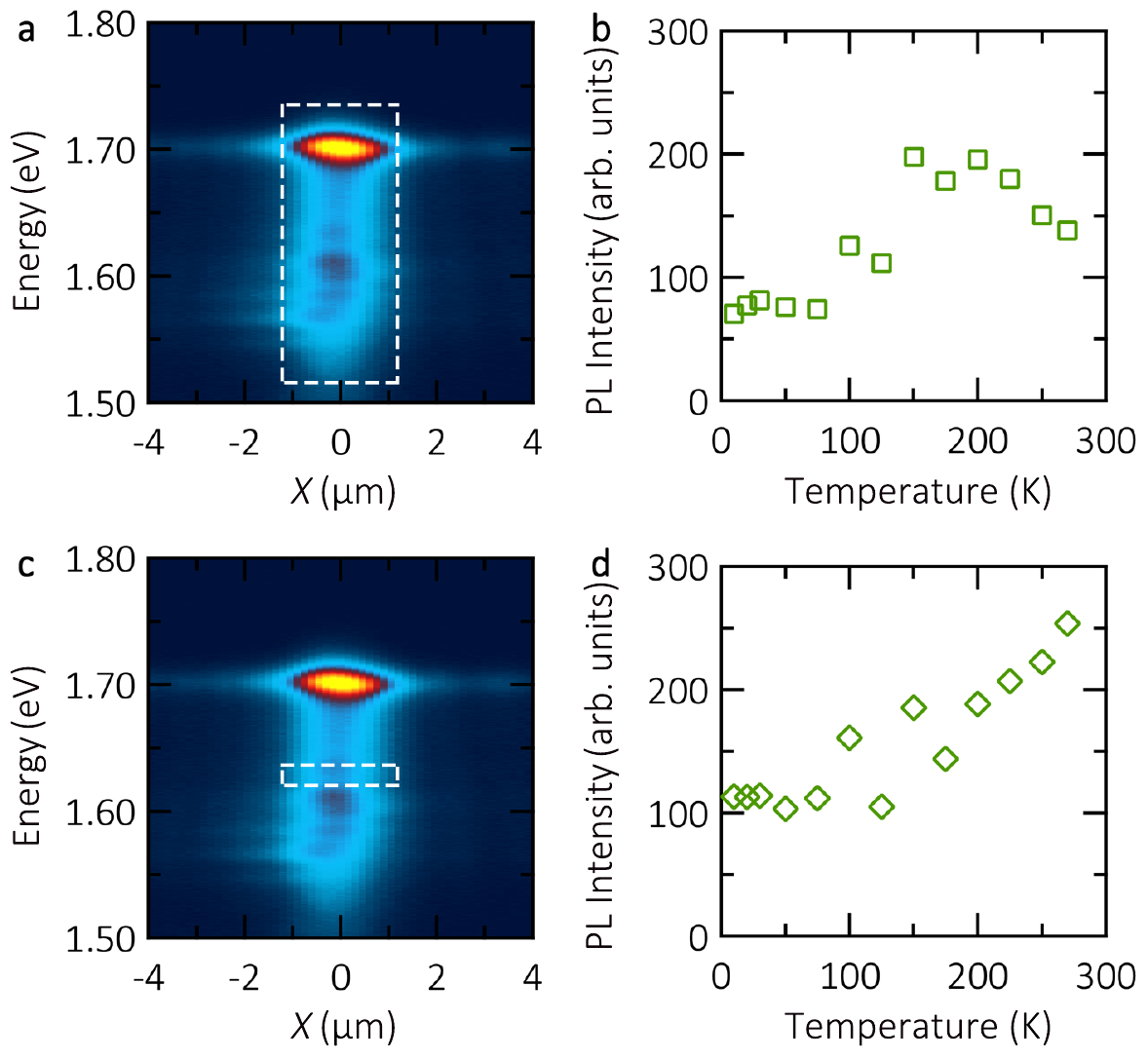}
\caption{\textbf {Integrated PL intensity of pristine $\rm WSe_2$ monolayer.} \textbf {a} is reproduced from Fig. 3a. The white box represents the integration region that ranges from 1.52 to 1.74 eV (including all emission features). \textbf {b} Temperature dependent emission intensity of pristine $\rm WSe_2$ monolayer, with the integration region shown in a. \textbf {c} The integration region ranges from 1.623 to 1.636 eV, corresponding to energies of ground state and first excited state of polaritons. \textbf {d} Temperature dependent emission intensity of pristine $\rm WSe_2$ monolayer, with the integration region shown in c.}\label{fig S9}
\end{figure}

%%\begin{figure}[h]%
%%\centering
%%\includegraphics[width=0.9\textwidth]{Fig S10.pdf}
%%\caption{\textbf {Polariton dispersion plotted in a larger photon energy scale.} \textbf {a} is reproduced from Fig. 3a. \textbf {b} It is reproduced from the data of Fig. 3d, plotted with the same photon energy scale as Fig. 3a.}\label{fig S10}
%%\end{figure}



\end{appendices}

\end{document}